\documentclass{jfm}

\usepackage{natbib}
\usepackage{amssymb,latexsym}
\usepackage{epstopdf}
\usepackage{graphicx,color}
\usepackage{psfrag}
\newcommand{\mylab}[3]{\raisebox{#2}[0mm][0mm]{%
\makebox[0mm][l]{\hspace*{#1}\textbf{#3}}}}
\def\spacce#1{\hskip #1pt}
\def\drawline#1#2{\raise 2.5pt\vbox{\hrule width #1pt height #2pt}}
\def\solid{\drawline{24}{.5}\nobreak}

\def\bdash{\hbox{\drawline{5.8}{.5}\spacce{2}}}

\def\dashed{\bdash\bdash\bdash\nobreak}

\def\chndot{\hbox%
{\drawline{4.6}{.5}\spacce{2}\drawline{1}{.5}\spacce{2}\drawline{4.6}{.5}\spacce{2}\drawline{1}{.5}\spacce{2}\drawline{4.6}{.5}}\nobreak }

\def\trian{\raise 1.25pt\hbox{$\scriptstyle\triangle$}\nobreak}

\def\dtrian{\raise 1.25pt\hbox%
{$\scriptscriptstyle\bigtriangledown$}\nobreak}

\def\squar{\raise 1.25pt\hbox{$\scriptstyle\Box$}\nobreak}

\def\diamon{\raise 1.25pt\hbox{$\scriptstyle\diamond$}\nobreak}

\newcommand{\soliddtrian}{$\blacktriangledown$\nobreak}

\def\linedtri1{\hbox{\bdash\hspace{-1.6mm}$\bigtriangleup$\hspace{-0.8mm}\bdash}\nobreak}
\def\soliddtrian1{$\blacktriangledown$\nobreak}
\def\solidrtrian2{$\blacktriangleright$\nobreak}
\def\solidltrian3{$\blacktriangleleft$\nobreak}
\def\dd{{\, \rm{d}}}

\def\bra{\langle}
\def\ket{\rangle}
\def\p{\partial}

\def\beq{\begin{equation}}
\def\eeq{\end{equation}}
\def\la{\label}
\def\ii{{\rm i}}

\def\eij{\varepsilon}
\def\hlambda{\widehat{\lambda}}
\def\hS{\widehat{S}}
\def\hD{\widehat{\Pi}}
\def\hu{\widehat{u}}

\def\hphi{\widehat{\phi}}

\def\lambdavec{\mbox{\boldmath $\lambda$}}
\def\phivec{\mbox{\boldmath $\phi$}}
\def\phimat{\mbox{\boldmath $\mathsfbi \Phi$}}
\def\Svec{\mbox{\boldmath $S$}}
\def\uvec{\mbox{\boldmath $u$}}
\def\utau{u_\tau}

\def\dis{\varepsilon}
\def\figpath{}
\def\r#1{(\ref{#1})}

\title[Optimal fluxes in turbulence]{Optimal fluxes and
Reynolds stresses}
\author{Javier Jim\'enez}
\affiliation{School of Aeronautics, Universidad Polit\'ecnica de Madrid, 28040 Madrid, Spain
}
\date{\today}	
\begin{document}
\maketitle											 

\begin{abstract}
It is remarked that fluxes in conservation laws, such as the Reynolds stresses in the
momentum equation of turbulent shear flows, or the spectral energy flux in isotropic
turbulence, are only defined up to an arbitrary solenoidal field. While this is not usually
significant for long-time averages, it becomes important when fluxes are modelled locally in
large-eddy simulations, or in the analysis of intermittency and cascades. As an example, a
numerical procedure is introduced to compute fluxes in scalar conservation equations in such
a way that their total integrated magnitude is minimised. The result is an irrotational
vector field that derives from a potential, thus minimising sterile flux `circuits'. The
algorithm is generalised to tensor fluxes and applied to the transfer of momentum in a
turbulent channel. The resulting instantaneous Reynolds stresses are compared with their
traditional expressions, and found to be substantially different.
\end{abstract}

\maketitle

\section{Introduction}\label{sec:intro}

Conservation laws are staples of continuum mechanics. They take the form of the rate of
change of a conserved quantity $\rho$, such as mass, energy or momentum density, balanced
by the divergence of a vector flux $\phivec=\{\phi_j\}$,
\beq
\p_t\rho +\p_j \phi_j = \widetilde{S},
\la{eq:conserv}
\eeq
where $\p_j$ is the partial derivative along the $j$-th coordinate, repeated indices imply
summation over all coordinate directions, and $\widetilde{S}$ represents any sources or
sinks. While the physical significance of the conserved quantity is usually obvious, that of
the flux is less clear, because only its divergence enters the equation. In
spite of this, the fluxes themselves are often given physical significance, such as when
Reynolds stresses are taken to represent the flux of momentum and are explicitly modelled in
large-eddy simulations, or when a constant energy transfer rate is used as the basic
parameter in the spectral theory of the turbulence cascade \citep{kol41}. The
cascade theories that form the backbone of modern turbulence research are not theories
about conserved quantities, but about their fluxes.
     
A consequence of these considerations is that fluxes cannot be uniquely defined.
Consider the generalisation of \r{eq:conserv},
\beq
\p_j \phi_j =\widetilde{S}-\p_t\rho \equiv S,
\la{eq:divn}
\eeq
where $S$ has been modified to include the instantaneous temporal rate of change of the densities,
and $\phivec$ is a vector flux in an $n$-dimensional space. In the first place, there is
often some ambiguity in which part of $S$ is incorporated into the flux, such as, for
example, when a constant pressure gradient $g$ is interpreted as a secular term $gx_j$
within the corresponding spatial flux. But, even if that decomposition is decided on
physical or other grounds, the definition of the fluxes remains ambiguous. The relation
\r{eq:divn} is singular when interpreted as an equation for the vector flux $\phivec$,
given the source $S$. For example, any solenoidal vector field can be added to $\phivec$ without
changing \r{eq:divn}, and a three-dimensional vector flux is only defined up to
the addition of a rotor.

This ambiguity is a particular case of the gauge invariance familiar from electromagnetic
and other field theories \citep{barut80}, and it should be clear that it in no way
invalidates the original field equations. As in the case of other classical field theories,
it only becomes important when trying to give physical significance to quantities that only
appear in the equations as a gradient or as a divergence. In those cases, gauge
transformations provides an extra degree of freedom in our choice of expression for the
fluxes that can be used to simplify further manipulations for specific purposes. However,
the extra gauge freedom implies that only gauge invariant quantities should be considered
to be physically relevant. For example, we will see below that the energy--momentum tensor
of classical fluid mechanics can be gauge-transformed. The same is true of its modelling
counterpart, the Reynolds or sub-grid stresses. This suggests that neither of them should be
a primary object of analysis, and that both should only be used as one among many possible
representations of the same physical object. Which representation to use in each particular
case should be decided on utilitarian, rather than on absolute grounds.

It should be noted that any two representations of the fluxes are linked by their
divergence. As a consequence, if a particular expression $R_j$ is known for the fluxes, there is
no need to compute the right-hand side of \r{eq:divn}. All other fluxes satisfy
\beq
\p_j \phi_j =\p_j R_j =S .
\la{eq:divnR}
\eeq

This paper describes a gauge designed to minimise a particular norm of $\phivec$. Although
these `optimal' fluxes have some useful properties, the emphasis is not so much on them as
on their comparison with fluxes defined in more classical ways. The main goal is to
determine whether different gauges result in very different flux properties, and how this
can be used to differentiate properties that are intrinsic to the physics from those linked to
a particular gauge. The paper is organised as follows. Optimal fluxes for scalar conservation
laws are introduced in \S\ref{sec:optimal}, and generalised to tensor fluxes of vector
equations in \S\ref{sec:tensor}. The methodology thus developed is applied to the stress
tensor of the momentum conservation equation for a turbulent channel in
\S\ref{sec:momentum}, and the results are compared to the classical Reynolds-stress tensor in
\S\ref{sec:results}. Conclusions and possible directions for future research are 
offered in \S\ref{sec:conclusions}.

\section{Optimal fluxes}\la{sec:optimal}

As an example of the previous considerations, we will develop expressions for a set of
`optimal' fluxes that minimise the integrated flux magnitude over a domain of interest
$\Omega$. It should be emphasized that this definition is not unique, and that it is only
optimum in the sense of minimising a particular norm. In general, choosing another norm or
even another domain results in a different expression, but we shall see that such fluxes
have sometimes a physically reasonable interpretation and that, as mentioned above,
comparing two alternative definitions may be useful to determine which properties of the
classically defined expressions are intrinsic to the physics, and which ones are accidents
of a particular choice of gauge. Define a cost function,
\beq
J = \int_\Omega [\phi_j\phi_j/2  +\lambda (\p_j\phi_j -S) ]\dd \Omega,
\la{eq:cost}
\eeq
where \r{eq:divn} has been added as a constraint with the scalar Lagrange multiplier
$\lambda$. Taking the first variation, $\phi_j \rightarrow \phi_j +\delta \phi_j$,
and integrating by parts, we obtain
\beq
\delta J = \int_\Omega (\phi_j  -\p_j \lambda)\, \delta \phi_j \dd \Omega + 
             \int_{\partial\Omega} \lambda\, \delta\phi_n \dd (\p\Omega) =0,
\la{eq:var1}
\eeq
where $\p\Omega$ is the boundary of $\Omega$, and  the `$n$' subscript denotes components normal to $\p\Omega$. Requiring  \r{eq:var1} to be satisfied for
arbitrary $\delta\phivec$ yields the Euler variational equations \citep{gelfand63},
\beq
\phi_j  =\p_j \lambda,
\la{eq:varphi}
\eeq
with natural boundary conditions, 
\beq
\lambda=0\quad\mbox{at}\quad \p\Omega.
\la{eq:varbc}
\eeq
The latter may require modification in especial cases. For example, if the fluxes are
assumed to be spatially periodic along some direction, $\lambda$ can also be assumed to be
periodic. Equation \r{eq:varphi} expresses the intuitive condition that the optimum flux should
be an (irrotational) gradient, `as free as possible' from circuits. When combined with the
dynamical relation \r{eq:divn}, the potential $\lambda$ satisfies the Poisson equation,
\beq
\nabla^2 \lambda=S,
\la{eq:pois}
\eeq
with homogeneous Dirichlet boundary conditions. Note that this does not imply that the
fluxes vanish at the boundary, but applying Gauss theorem to \r{eq:divnR} shows that
\beq
\int_{\p\Omega} \phi_n \dd(\p\Omega)=\int_{\p\Omega} R_n \dd(\p\Omega).
\la{eq:totflux}
\eeq
The total flux across the boundary is independent of the representation.

%
\subsection{Tensor fluxes}\la{sec:tensor}

Equation \r{eq:divn} can be generalised to a vector right-hand side $\Svec=\{S_i\}$, such
as momentum, and to a tensor flux $\phimat = \{\phi_{ij}\}$,
\beq
\p_j \phi_{ij} =S_i.
\la{eq:divn2}
\eeq
The potential is then a vector $\lambdavec=\{\lambda_i\}$, and the problem can be
defined as minimising the integrated Euclidean norm of the tensor. However, it is often
the case that $\phimat$ is not fully arbitrary, and the minimisation has to
consider additional constraints. For example, the momentum flux tensor should be
symmetric, in which case the cost function is
\beq
J = \int_\Omega [\phi_{ij}\phi_{ij}/2  +\lambda_i (\p_j\phi_{ij} -S_i) +\eij_{mij}\mu_m\phi_{ij}]
\dd \Omega,
\la{eq:costphi}
\eeq
where $\eij_{mij}$ is the completely antisymmetric Levi--Civita symbol, and the 
$\mu_m$ are extra Lagrange multipliers to ensure the symmetry of $\phi_{ij}$. The
Euler equations are then
\beq
\phi_{ij}=\p_j\lambda_i-\eij_{mij}\mu_m,
\la{eq:varphij}
\eeq
with natural boundary conditions as in \r{eq:varbc}. The requirement that $\phi_{ij}
=\phi_{ji}$ implies
\beq
\eij_{mij}\mu_m =(\p_j\lambda_i-\p_i\lambda_j)/2,
\la{eq:mum}
\eeq
and \r{eq:varphij} becomes
\beq
\phi_{ij} =(\p_j\lambda_i+\p_i\lambda_j)/2.
\la{eq:varphisym}
\eeq
Substituting in \r{eq:divn2} results in 
\beq
\nabla(\nabla\cdot\lambdavec)+ \nabla^2 \lambdavec = 
2\nabla(\nabla\cdot\lambdavec)- \nabla\times \nabla\times \lambdavec = 
2\Svec.
\la{eq:poisvec}
\eeq
This vector equation proves that $\lambdavec$ is a cartesian vector, and that $\phimat$ is a
cartesian tensor. A useful equation for the trace of $\phimat$,
$\Pi=\phi_{ii}=\nabla\cdot\lambdavec$, is obtained taking the divergence of \r{eq:poisvec},
\begin{eqnarray}
\nabla^2\Pi &=& \nabla\cdot\Svec.
\la{eq:poisdiv}%
\end{eqnarray}
If desired, $\Pi$ can be separated from $\phimat$ as a pressure-like isotropic term, and
\r{eq:poisdiv} becomes a variant of the usual pressure equation. In fact, the procedure
leading to \r{eq:varphi}--\r{eq:pois} is akin to the classical derivation of the pressure
equation in incompressible flows, and $\phimat$ can be loosely interpreted as a generalised
`tensor pressure' that completes the right-hand side of \r{eq:divn2} in the same sense that
the standard scalar pressure projects the momentum equation onto the incompressible
subspace. On the other hand, the interpretation of the optimal tensor flux is not as
straightforward as for a vector, since there is nothing like a rotor to justify the
interpretation of `minimum circularity'. The condition of minimum magnitude remains.

\section{Momentum transfer in a turbulent channel}\label{sec:momentum}

We illustrate the above procedure by computing the optimal momentum fluxes in a
pressure-driven incompressible turbulent channel between infinite parallel plates separated
by $2h$. As mentioned above, our main purpose is to determine how different are the optimal
momentum fluxes from the classical Reynolds stresses, and thereby which properties of
the latter should be considered physical and which ones accidental. We denote by $x_i$, with
$i=1$ to 3, the streamwise, wall-normal and spanwise coordinates, respectively, with $x_2=0$
at the lower wall. Momentum is injected across the channel cross-section by the streamwise
gradient $g_1$ of the kinematic pressure, uniform in space but not necessarily in time. The
spanwise pressure gradient vanishes at all times. Momentum is removed at the walls by
viscous friction, and has to be transferred along $x_2$ from the body of the flow to the
wall. The resulting flux is conserved except for the constant pressure forcing, and its
conservation is responsible for the possibility of using the friction velocity $\utau$ as a
uniform velocity scale at all wall distances \citep{towns}. Quantities normalised with
$\utau$ and with the kinematic viscosity $\nu$ are denoted by a `+' superscript.

The structures responsible for this transfer have been studied often. A recent summary can
be found in \cite{lozano-Q}, where it is shown that three-dimensional `quadrant' structures
\citep{wall:eck:bro:72,lu:wil:73} exist at all scales, and that the most intense ones form a
self-similar hierarchy of sweeps and ejections with sizes proportional to their distance to
the wall. Because of this size stratification, it can be argued that the momentum transfer
constitutes an inertial turbulent cascade taking place mostly in space
\citep{jim12_arfm,jim13_rev}, although different from the energy cascade in \cite{kol41}.
However, the non-uniqueness of the fluxes raises the question of the generality of these
structures and of their properties.

The momentum equation can be written as
\beq
 \p_j \phi_{ij}  = -\p_t u_i-g_i,
\la{eq:NS1}
\eeq
which is satisfied by 
\beq
\phi_{ij}=R_{ij} \equiv u_i u_j +p\delta_{ij}-2\nu\sigma_{ij},
\la{eq:NS1bis}
\eeq
where $\sigma_{ij}=(\p_i u_j+\p_j u_i)/2$ is the rate-of-strain tensor, and $\delta_{ij}$ is
Kronecker's delta. The left-hand side of  \r{eq:NS1} has been written in the form of a
divergence, but we will see below by direct calculation that the tensor flux
$R_{ij}$ is not optimal. We are not aware of any analytic expression for the optimal flux
tensor $\phi_{ij}$ associated with \r{eq:NS1}, but the algorithm discussed above can be
easily implemented numerically, and our interest will be to explore how the instantaneous
optimum fluxes for \r{eq:NS1} differ from their classical analytic expressions. We will be
particularly interested in the tangential flux $\phi_{12}$, which is the only one that
survives under long-time averaging. Introducing $\bra\ket$ to denote averaging over
wall-parallel planes and time, it follows from \r{eq:NS1} that
\beq
\p_2 \bra\phi_{12}\ket  + \bra g_1\ket =0.
\la{eq:phi12}
\eeq
Note that the pressure gradient $g_1$ has been taken outside the divergence in \r{eq:NS1},
because of its obvious physical interpretation as a momentum source. This also allows us to
define quantities that are periodic in $x_1$, including the residual pressure $p$ and the
diagonal momentum fluxes $\phi_{jj}$ (no summation).

We use simulations in a doubly periodic channel with reasonably large streamwise and
spanwise dimensions $L_1\times L_3 = 8\pi h\times 3\pi h$, and $h^+=934$ \citep{juanc04}.
The different variables can then be expanded in Fourier series as in
\beq
u_i  = \hu_{i,\alpha}(x_2) \exp[\ii(\alpha_1 x_1 + \alpha_3 x_3)],
\la{eq:ufou}
\eeq
where $\alpha_k(j) = 2\pi j/L_k,\, j=-\infty\ldots-1,0,1\ldots\infty$, for $k=1$ or $k=3$.
For the rest of the paper, the dependence of the Fourier coefficients on the wavenumber
will be omitted. Consider, for example, the
streamwise component of the flux potential equation \r{eq:poisvec},
\begin{eqnarray}
\p_{22}\hlambda_{1} -(\alpha_1^2+\alpha_3^2) 
\hlambda_{1}  &=& 2\hS_{1} - \ii\alpha_1 \hD, 
\la{eq:poisF1}
\\
\hlambda_{1}(0) = \hlambda_{1}(2h) &=& 0, 
\la{eq:poisF2}
\end{eqnarray}
where $\hS_i$ is the right-hand side of \r{eq:NS1}, and 
\beq
\hD = \widehat{\nabla\cdot \lambdavec} = 
\ii\alpha_1\hlambda_{1}+\p_2\hlambda_{2} + \ii\alpha_3\hlambda_{3},
\la{eq:divfou}
\eeq
The fluxes become,
\begin{eqnarray}
\hphi_{11} &=& \ii\alpha_1 \hlambda_{1},
\la{eq:poisF3}
\\
\hphi_{12} = \hphi_{21} &=& (\p_2\hlambda_{1}+\ii \alpha_1\hlambda_{2})/2,
\la{eq:poisF4}
\\
\hphi_{13} = \hphi_{31} &=& \ii (\alpha_3 \hlambda_{1}+\alpha_1 \hlambda_{3})/2.
\la{eq:poisF5}
\end{eqnarray}
The divergence $\hD$ satisfies Poisson's equation \r{eq:poisdiv} with unknown boundary
conditions. These are handled indirectly, as in the channel simulations of \cite{kmm}.
The equation for each Fourier component of the divergence is solved three times: one with
its full right-hand side and homogeneous boundary conditions,
\beq
\p_{22}\hD_0 -(\alpha_1^2+\alpha_3^2) \hD_0  = \widehat{\nabla\cdot \Svec}, 
\qquad
\hD_0(0) = \hD_0(2h) = 0, 
\la{eq:poisdivF1}
\eeq
and two with a homogeneous right-hand side and unit boundary condition at one wall
and zero at the other. For the solution associated with the lower wall, 
\beq
\p_{22}\hD_L -(\alpha_1^2+\alpha_3^2) \hD_L  = 0, 
\qquad
\hD_L(0) =1,\quad \hD_L(2h) = 0, 
\la{eq:poisdivF2}
\eeq
with an equivalent expression for the upper one, $\hD_U$. The divergence can then be written as
$\hD=\hD_0 +a_L\hD_L + a_U \hD_U$, which satisfies,
\beq
\hD(0) =a_L,\qquad \hD(2h) = a_U. 
\la{eq:divBC}
\eeq
The Poisson problem \r{eq:poisF1}--\r{eq:poisF2} is solved three times for each
$\lambda_i$. For example, once for $\hlambda_{10}$, with right-hand side $\hS_1-\ii\alpha_1
\hD_0$, and once for each of $\hlambda_{1L}$ and $\hlambda_{1U}$, with right-hand sides
$-\ii\alpha_1 \hD_L$ and $-\ii\alpha_1 \hD_U$, respectively. Again, the solution can be
expressed as $\hlambda_{1}=\hlambda_{10} +a_L\hlambda_{1L} + a_U \hlambda_{1U}$. The
process is repeated for $\hlambda_2$ and $\hlambda_3$, allowing us to compute the
divergence of the resulting $\lambdavec$ from its definition \r{eq:divfou}. The values of
this divergence at the two walls are also linear combinations of three terms, two of which
are proportional to $a_U$ and $a_L$. Substituting them in \r{eq:divBC} allows these
coefficients to be computed, and the problem to be closed.

\subsection{Fluctuation velocities}\la{sec:fluctuations}

The different components of the fluxes $R_{ij}$ in \r{eq:NS1bis} have very dissimilar
magnitudes. Equation \r{eq:phi12} can be integrated to give $\bra \phi_{12}^+\ket =\bra
\phi_{12}\ket/\utau^2 =1-x_2/h$, where $\utau^2=-\bra g_1\ket$. This is satisfied by all
flux definitions, and implies that the mean of $\phi_{12}$ is $O(1)$ in wall units. However,
the classical fluxes given by \r{eq:NS1bis} are found to have standard deviations of order
$R_{rms}^+=O(20-60)$, which can be used as proxies for their integrated euclidean norm. The
optimisation procedure reduces these intensities considerably (not shown), but only to
$\phi_{rms}^+=O(10)$. This is important because, if instantaneous fluxes are to be used to
study their contribution to the mean momentum transport or to model them in LES, it is
useful if their characteristic magnitude is not much larger than their average.

Some reflection shows that the problem is that \r{eq:NS1}--\r{eq:NS1bis} are written in the frame of
reference linked to the wall, and that the fluctuating fluxes are dominated by sweeping
terms of the type $U u'_i$, where we have made the customary decomposition,
$u_i=U(x_2)\delta_{i1}+u_i'$ with respect to the mean profile $U(x_2) = \bra u_1\ket$.
In terms of the perturbation velocities, the momentum equation becomes,
\beq
\p_j \phi'_{ij} = -\p_t u'_i+g_i -U\p_1u'_i +(\nu \p_{22} U -u'_2\p_2U)\delta_{i1},
\la{eq:NS2}
\eeq
one of whose solutions is
\beq
\phi'_{ij}=R'_{ij} \equiv u'_i u'_j +p\delta_{ij}-2\nu\sigma'_{ij}, 
\la{eq:NS2bis}
\eeq
with $\sigma'_{ij}$ defined as in \r{eq:NS1} using $\uvec'$. When compared with \r{eq:NS1},
most of the extra terms in the right-hand side of \r{eq:NS2} average to zero over long
times, but they can be large instantaneously, and are responsible for the large standard
deviations of the fluxes in \r{eq:NS1bis}. For example, it is known experimentally that the standard
deviation of the perturbation tangential stress in the logarithmic layer is $(u'_1
u'_2)_{rms}^+\approx 2$ \citep{lozano-Q}, but the standard deviation
of $(u_1 u_2)^+$ is $O(20)$. There is no difference between perturbation and total
velocities for the transverse velocity components.

The fluctuation intensities of the perturbation fluxes in \r{eq:NS2bis} are
represented in figure \ref{fig:flux2} as lines with symbols. They are weaker than the values
cited above for the fluxes based on the full velocities, and only the standard deviation of
$R^{\prime +}_{11}$ reaches $O(10)$ near the wall. However, it should be born in mind that
these fluxes no longer represent the full momentum transfer, and that part of the
momentum is now carried by the linear advection terms in the right-hand side of \r{eq:NS2}.
The left-most one, $U\p_1 u'_1$, is the advection of the velocity fluctuations by the mean
velocity profile, and appears as flux fluctuations in any but the semilagrangian frame of
reference that follows the mean flow. It was shown in \cite{jim13_lin} that about 90\% of
the particle acceleration in a channel flow is due to this term, and this is the main reason
why the standard deviation of the fluxes is reduced when defined in terms of fluctuations.
The price of this nonuniform frame of reference is the last term in the right-hand side of
\r{eq:NS2}, $u'_2\p_2 U$, which is the classical lift-up representing the change in mean
velocity of a fluid particle as it moves with respect to the wall. Whether these transfers
should be treated as fluxes or sources has to be decided on physical grounds.

\section{Results}\la{sec:results}

\begin{figure}
\centerline{%
\includegraphics[width=0.47\textwidth]{\figpath 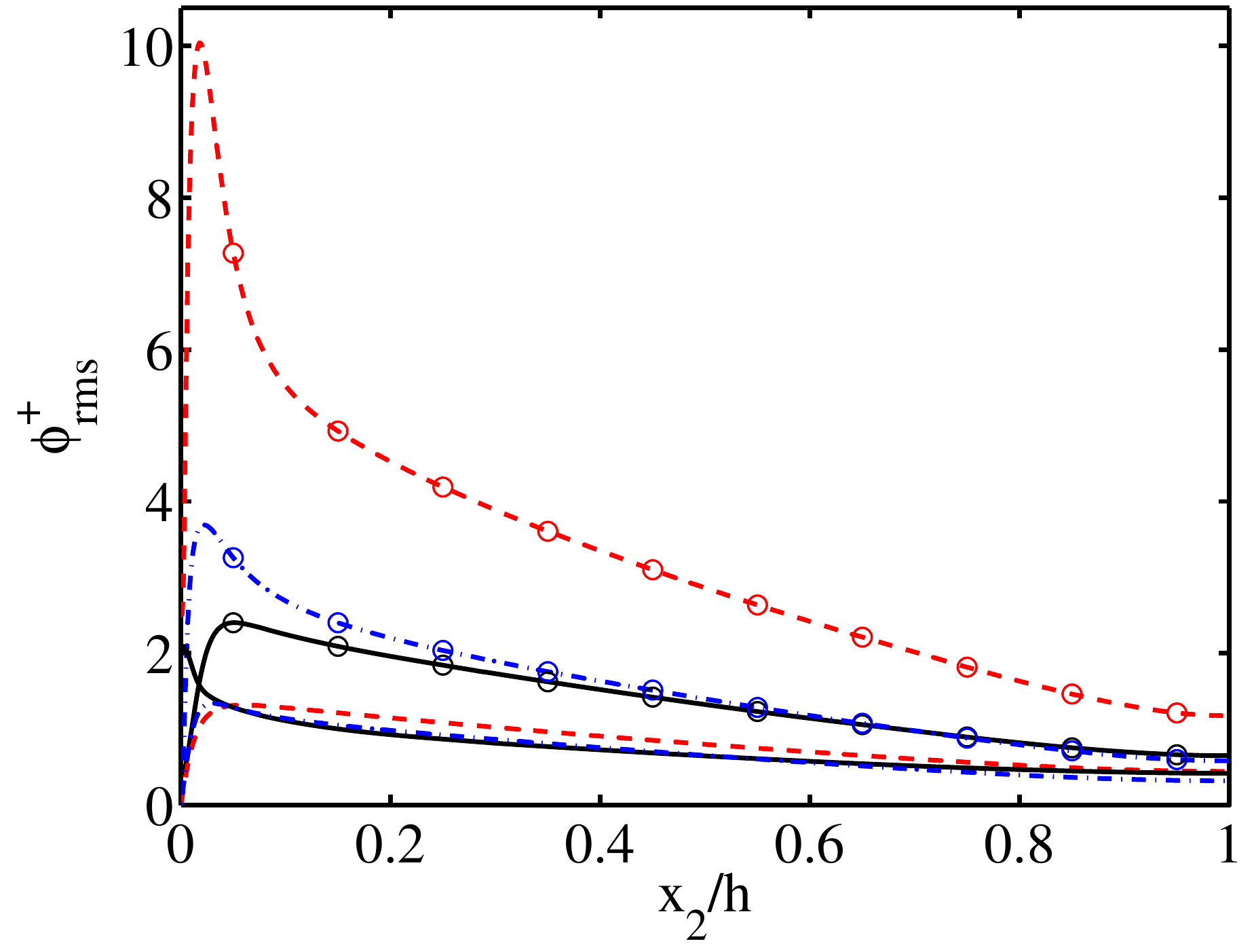}%
\mylab{-0.07\textwidth}{0.30\textwidth}{(a)}%
\hspace{1mm}%
\includegraphics[width=0.47\textwidth]{\figpath 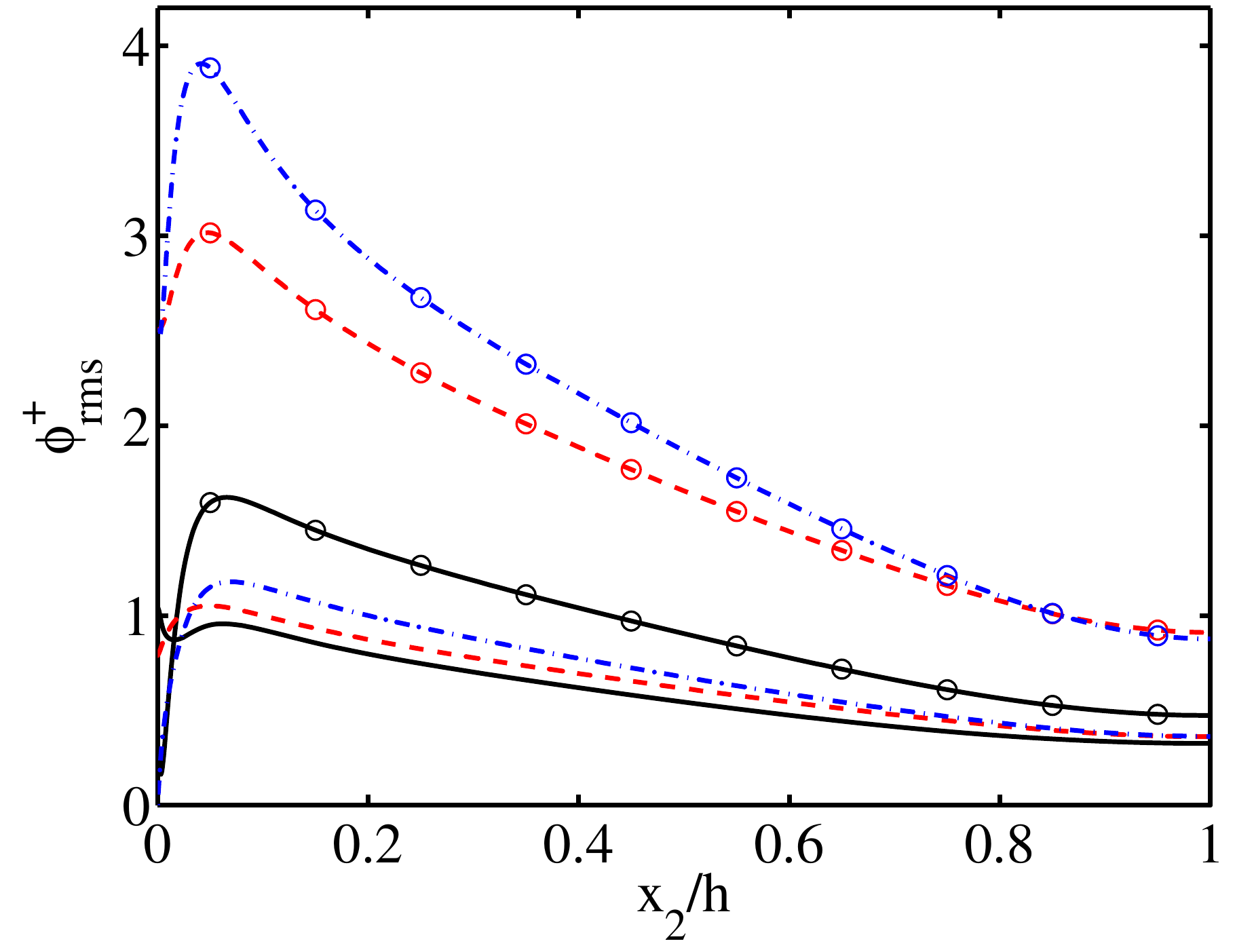}%
\mylab{-0.07\textwidth}{0.30\textwidth}{(b)}%
}
\caption{Root-mean-squared fluctuation intensities of the momentum fluxes along the three
coordinate directions, computed with the fluctuation velocity equation \r{eq:NS2}.
(a) Fluxes of the streamwise momentum. \dashed, $\phi_{11}$; \solid, $\phi_{12}$; \chndot,
$\phi_{13}$.
(b) Fluxes of the transverse momenta. \dashed, $\phi_{22}$; \solid, $\phi_{23}$; \chndot,
$\phi_{33}$.
Lines with circles are $R'_{ij}$ from \r{eq:NS2}. Those without symbols are optimum fluxes $\phi'_{ij}$
from (\ref{eq:poisF1}--\ref{eq:poisF5}).
}
\label{fig:flux2}
\end{figure} 

The results of applying the optimisation process to \r{eq:NS2} are shown in figure
\ref{fig:flux2}, where they are compared with the classical algebraic perturbation fluxes
$R'_{ij}$. The optimisation reduces the fluctuating intensity of all the fluxes by a
substantial factor that varies among components. In fact, the reduction is larger than
shown in the figure, because the standard deviation is computed with respect to the mean
value, which is typically not zero for the classical fluxes. For example, $\bra R'_{11}\ket
= \bra {u'_1}^2\ket\ge 0$, while it follows from \r{eq:varphisym} that the mean value of the
optimal diagonal fluxes along any homogeneous direction vanishes identically. All the
optimal flux fluctuations are of similar magnitude, and of the order of the mean momentum
transfer rate $\utau^2$. Note that the standard deviations discussed here refer to the
fluctuations of the quadratic functions of the velocities, as in $(u^2)_{rms}^2= \bra
u^4\ket -\bra u^2\ket^2$, not to those of the velocities themselves.

Although not shown in the figure, the effect of the pressure on the diagonal stresses
$R'_{ii}$ in \r{eq:NS2bis} (no summation implied) is not negligible, and always increases the
fluctuation intensities when compared with the classical Reynolds products ${u'_i}^2$. This
is particularly evident for the wall-normal velocity $u_2^2$. On the other hand, the effect of
the viscous term in $R'_{ij}$ is negligible above $x_2^+\approx 20$.

\begin{figure}
\centerline{%
\includegraphics[width=0.47\textwidth]{\figpath 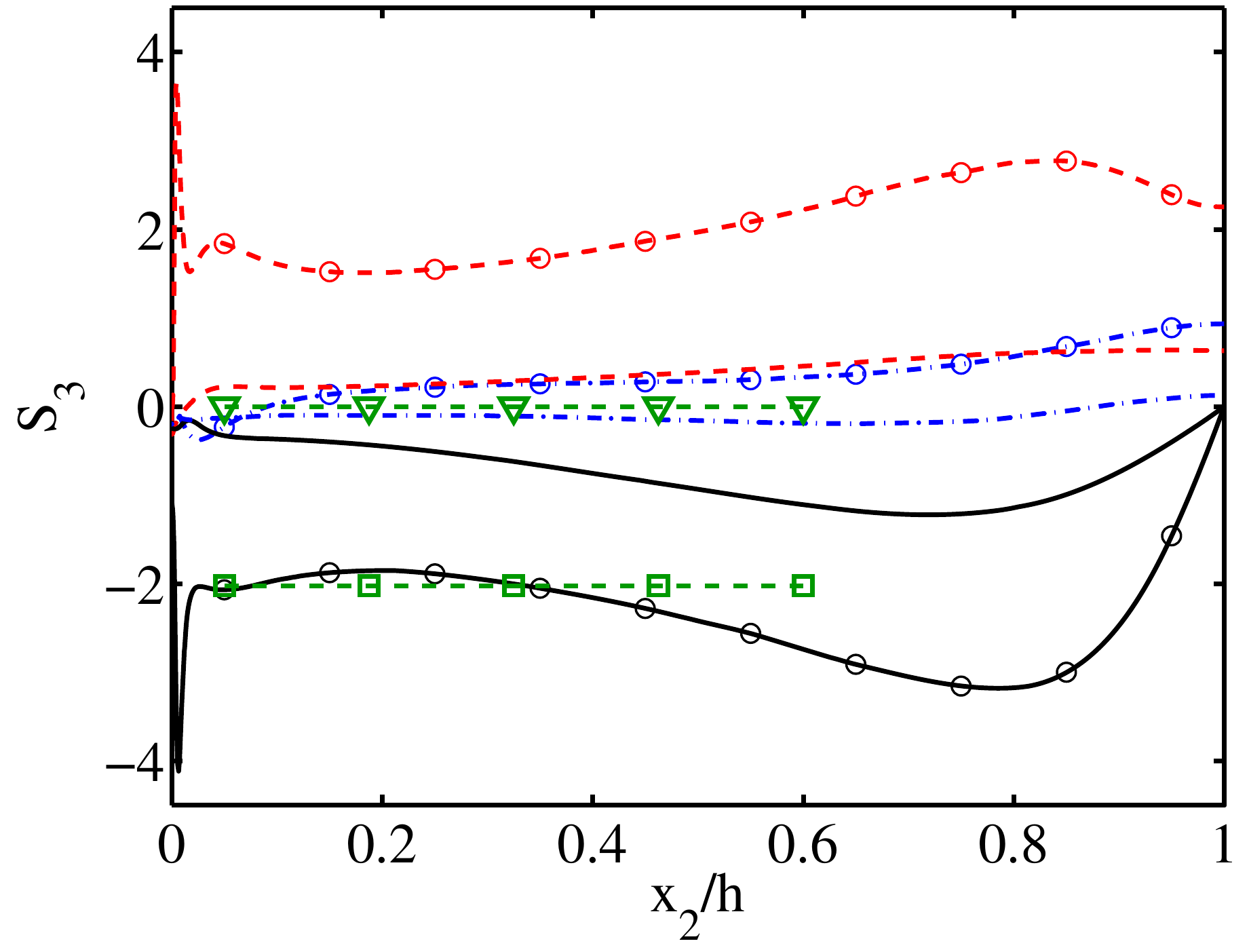}%
\mylab{-0.37\textwidth}{0.32\textwidth}{(a)}%
\hspace{1mm}%
\includegraphics[width=0.47\textwidth]{\figpath 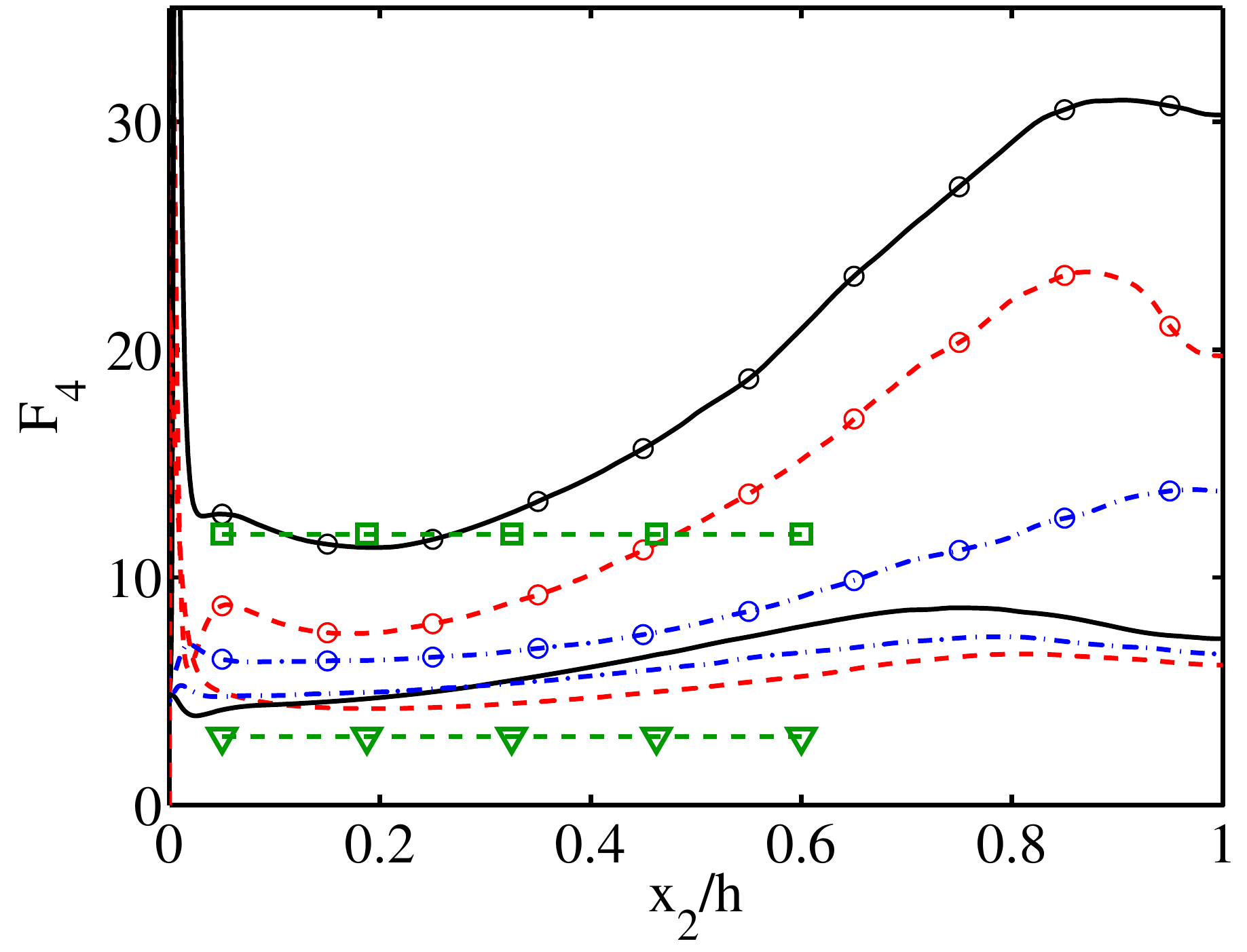}%
\mylab{-0.37\textwidth}{0.32\textwidth}{(b)}%
}%
\vspace{2mm}%
\centerline{%
\includegraphics[width=0.47\textwidth]{\figpath 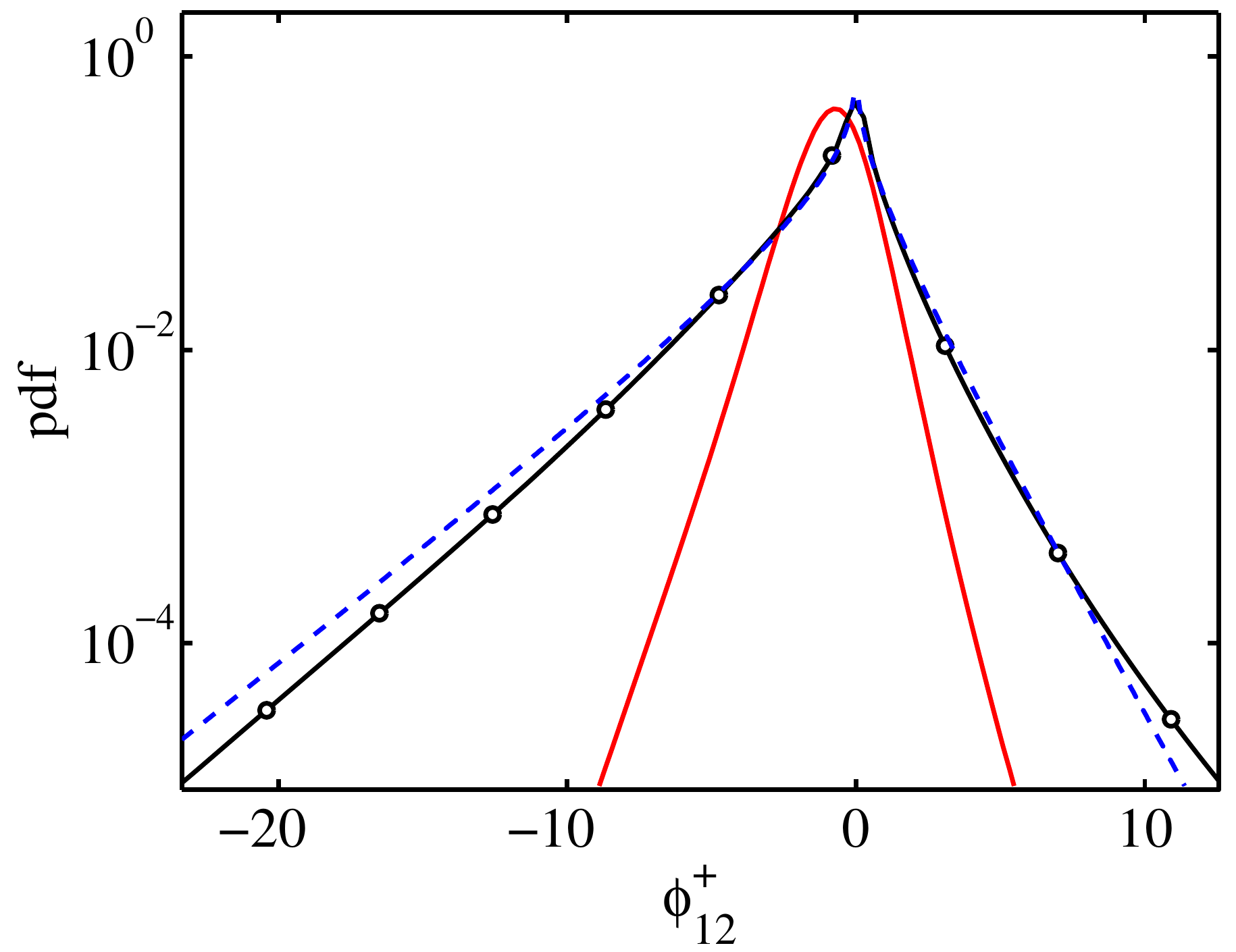}%
\mylab{-0.37\textwidth}{0.32\textwidth}{(c)}%
\hspace{1mm}%
\includegraphics[width=0.47\textwidth]{\figpath 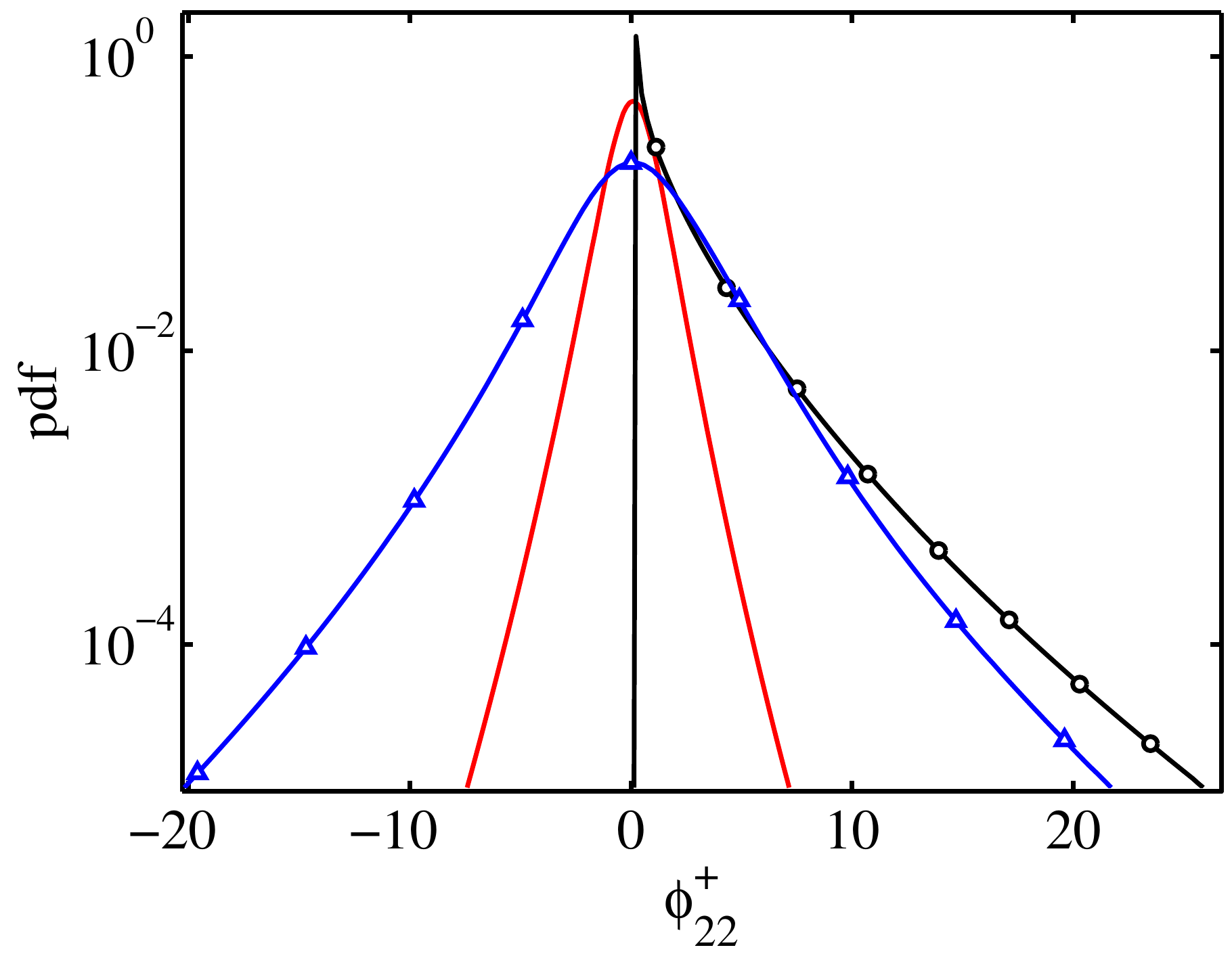}%
\mylab{-0.37\textwidth}{0.32\textwidth}{(d)}%
}
\caption{Higher-order moments of the fluctuation intensities of the centred momentum
fluxes, computed from the fluctuation velocities. \dashed, $\phi'_{11}$; \solid,
$\phi'_{12}$; \chndot, $\phi'_{22}$.
(a) Third-order skewness, $\bra \phi'^3\ket/ \bra \phi'^2\ket^{3/2}$. 
(b) Fourth-order flatness. $\bra \phi'^4\ket/ \bra \phi'^2\ket^{2}$.
The horizontal dashed lines are theoretical values for different functions of
gaussian-distributed variables. \dtrian, A gaussian variable $(S_3=0,\,F_4=3)$; \squar, the
product of two gaussian variables with cross-correlation coefficient $-0.4$
$(S_3=-2.02,\,F_4=11.9)$.
(c) Pdf of $\phi_{12}$ at $x_2/h\in(0.1-0.2)$, normalised in wall units. The dashed line is
the product of two gaussian variables with cross-correlation coefficient $-0.4$.
(d) As in (c), for $\phi_{22}$. The line with triangles is for $u'^2_2+p$.
In all figures, lines with circles are the Reynolds products, $u'_i u'_j$, and those without symbols
are optimum fluxes from (\ref{eq:poisF1}--\ref{eq:poisF5}).
}
\label{fig:interm}
\end{figure} 

The optimal fluxes are also less intermittent than the classical algebraic ones or than the
Reynolds products $\tau_{ij}=u'_iu'_j$. Their third-order skewness and fourth-order flatness are given
in figure \ref{fig:interm}(a,b). It is well-known that the Reynolds products are skewed and
intermittent, which is clear from the figure, but this is mostly a consequence of their
definition as quadratic forms. For example, even if a variable is gaussianly distributed,
its square is not, and \cite{AntAtk73} and \cite{lu:wil:73} showed that the probability
distributions of the product $u'_1 u'_2$ is essentially the same as the product of two
gaussian variables with the correct cross-correlation coefficient. The theoretical moments
for this product of gaussian variables are given in figure \ref{fig:interm}(a,b), and
represent well the observations for the tangential Reynolds product, except very near and
far from the wall. The optimal fluxes, which do not suffer from these algebraic artefacts,
are much less intermittent and stay approximately gaussian except in the buffer layer.
Although not shown in the figure to avoid clutter, the effect of the pressure on $R'_{ij}$
is to decrease intermittency. Particularly for $u'^2_2$ and $u'^2_3$, the flatness of the
Reynolds products is about three times higher than for the corresponding $R'_{ii}$, but the
effect of the viscous term is also negligible in this case. 

The probability density functions (pdfs) for two flux components in the `logarithmic' layer are
given in figure \ref{fig:interm}(c,d), where both the smaller standard deviation and the
weaker intermittency are clear. In the case of $\tau_{12}$ in figure \ref{fig:interm}(c),
the figure also shows the theoretical pdf for the product of two gaussian variables, which
fits the classical Reynolds product well except at the extreme tails. Note that the mean
value of these two pdfs should be exactly the same, $\bra \phi_{12}^+\ket =x_2/h-1$, but is
achieved by the two fluxes in different ways, While the pdf of the classical Reynolds
product peaks at $\tau_{12}=0$, and owes its negative mean value to the skewness of its
tails, the distribution of the optimal fluxes is roughly symmetric about its negative
mean value.

Figure \ref{fig:interm}(d) displays the pdf of the diagonal stress $\phi_{22}$. It also
shows the narrower distribution of the optimal flux and its narrower tails, although the
main difference in this case is that the classical product, $u'^2_2$, is intrinsically
positive. The figure also shows the effect of the pressure, discussed at the beginning of
this section. Its main effect is to restore the approximate symmetry of the pdf of
$u'^2_2+p$, which now includes negative values. As mentioned above, this also decreases
intermittency, although figure \ref{fig:interm}(d) shows that most of this decrease is due
to the effect of centring the one-sided pdf of the square. One the other hand, there almost
no difference between the pdf of the optimum $\phi'_{22}$ and that of its traceless
equivalent, $\phi'_{22}-\Pi'/3$ (not shown).

\begin{figure}
\centerline{%
\begin{minipage}[b]{0.80\textwidth}%
\vspace{6mm}%
\includegraphics[width=\textwidth]{\figpath 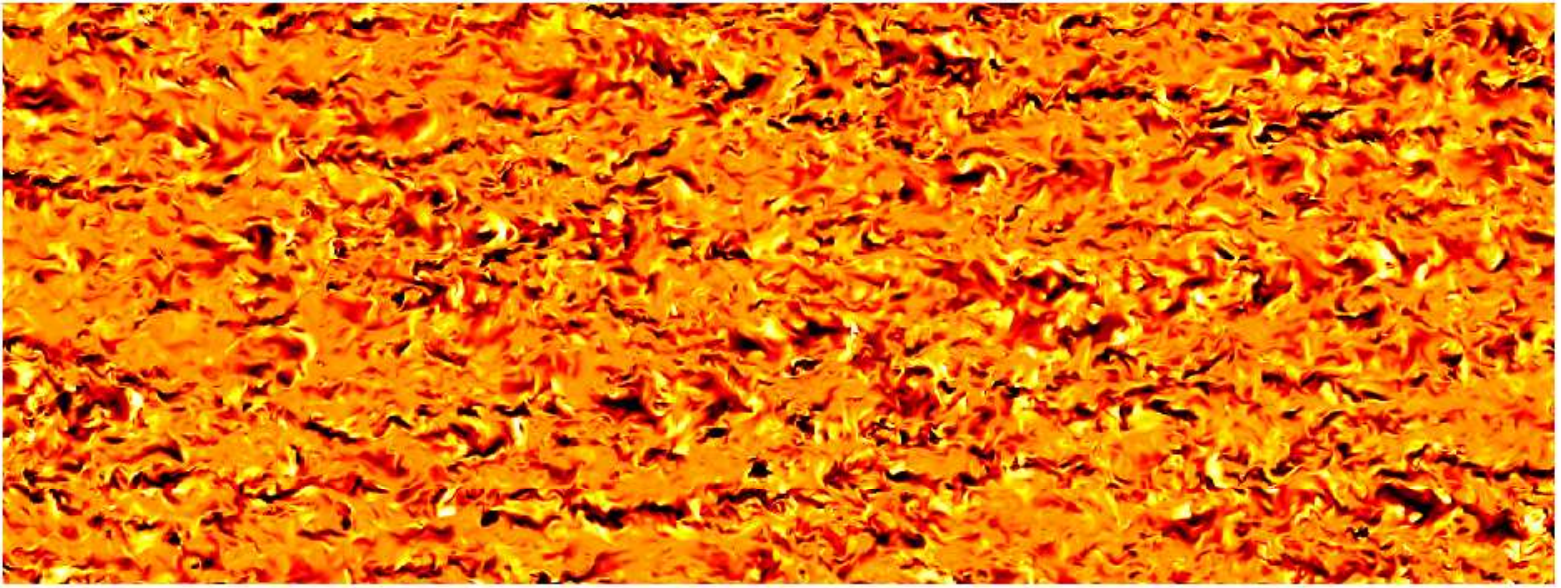}%
\mylab{-0.51\textwidth}{0.40\textwidth}{(a)}%
\\[6mm]%
\includegraphics[width=\textwidth]{\figpath 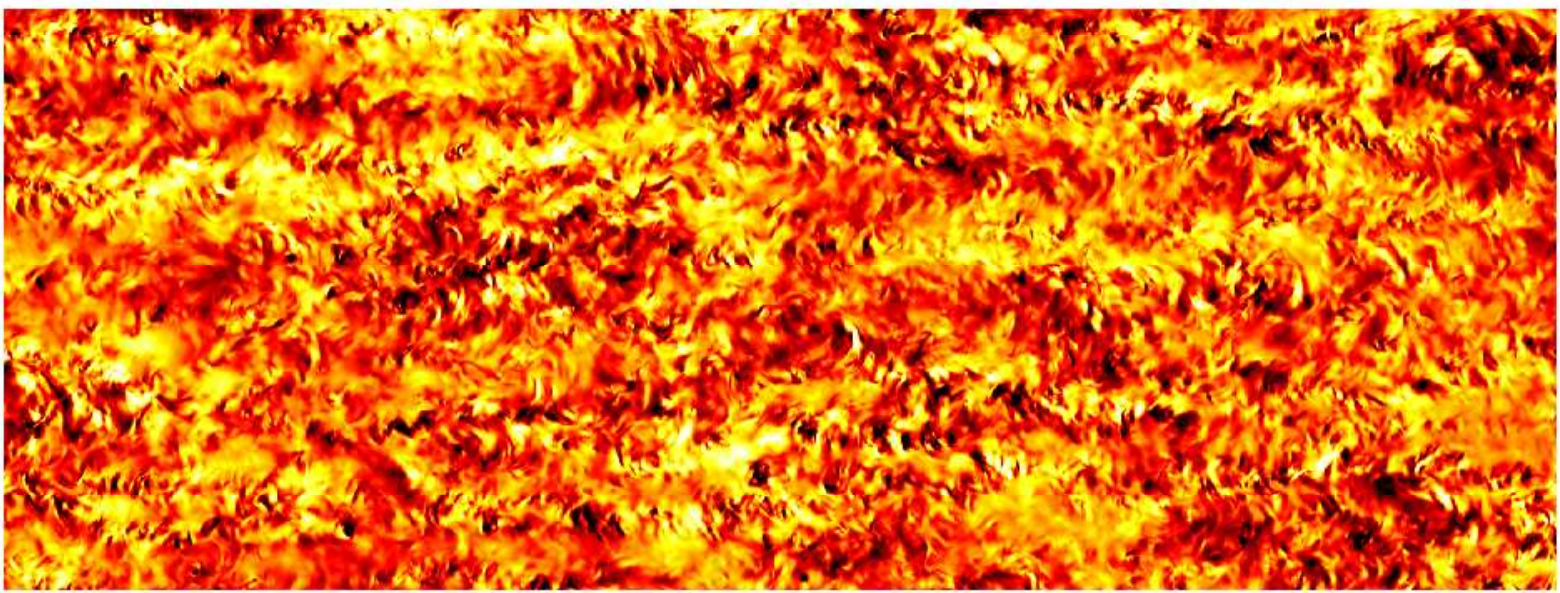}%
\mylab{-0.51\textwidth}{0.40\textwidth}{(b)}%
\end{minipage}%
}
\caption{%
Wall-parallel snapshots of the instantaneous tangential momentum flux in a channel at
$Re_\tau=934$ \citep{juanc04} and $x_2/h=0.15$. Flow is from left to right, and the area in
the figures is $L_1\times L_3 =4\pi h\times 3\pi h/2$. (a) Classical flux $R'_{12}$ from
\r{eq:NS2bis}. (b) Optimal flux $\phi'_{12}$. In both cases, the flux is centred with its
mean value, and the colour scale spans $\pm 3$ standard deviations, increasing from dark to
light.
}
\label{fig:flux3}
\end{figure} 

\begin{figure}
\centerline{%
\includegraphics[width=0.47\textwidth]{\figpath 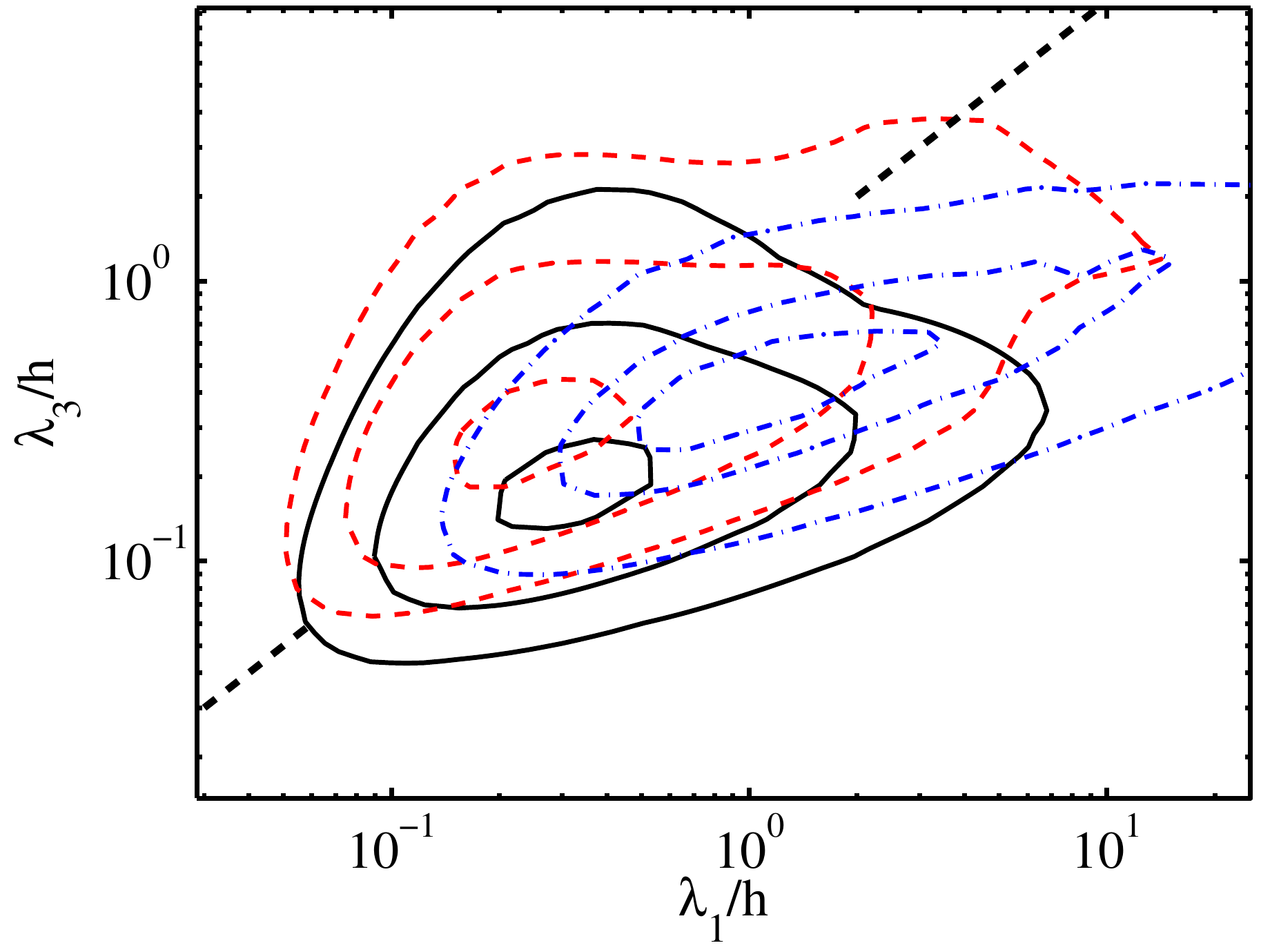}%
\mylab{-0.37\textwidth}{0.31\textwidth}{(a)}%
\hspace{2mm}%
\includegraphics[width=0.465\textwidth]{\figpath 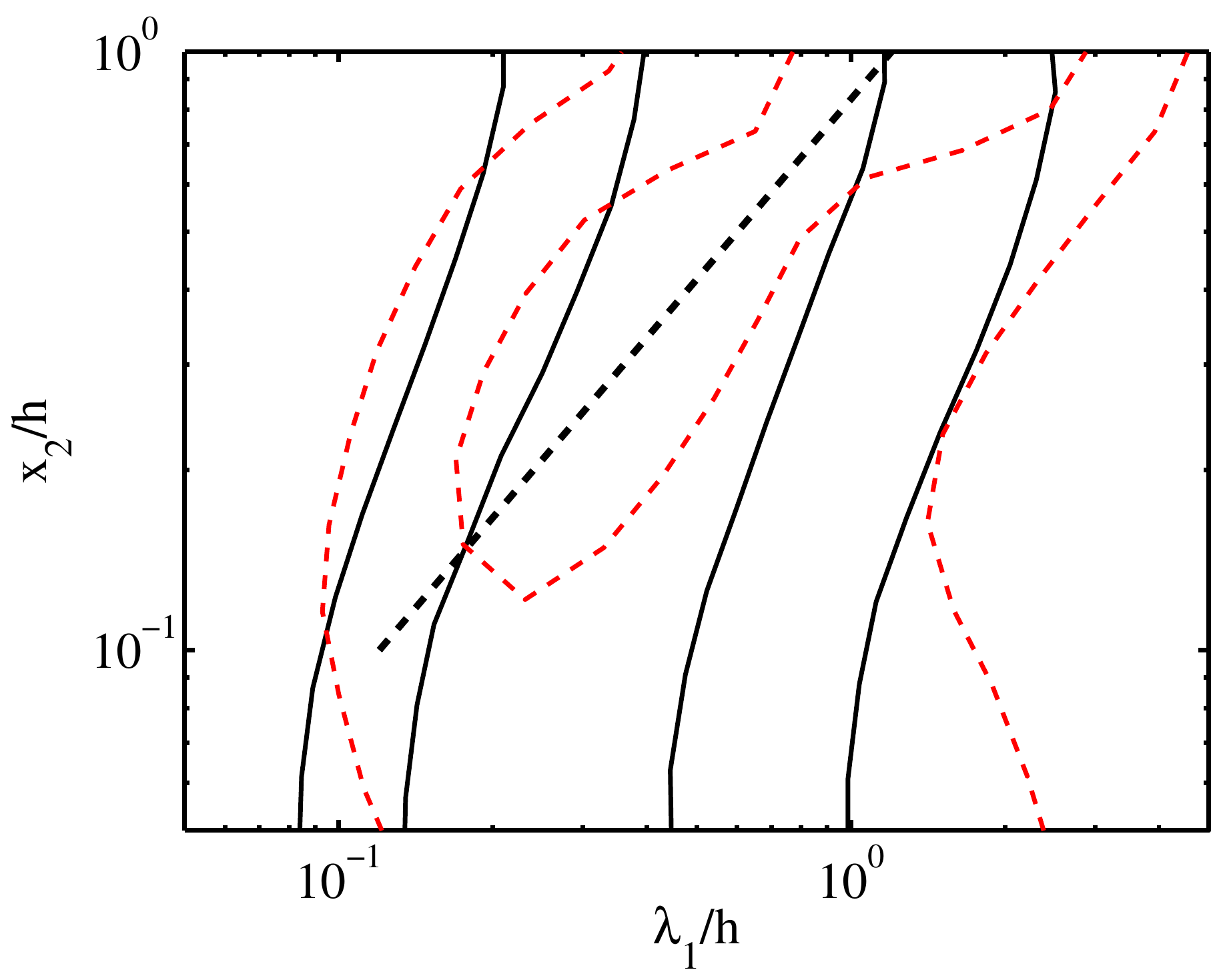}%
\mylab{-0.37\textwidth}{0.31\textwidth}{(b)}%
}
\caption{%
(a) Two-dimensional premultiplied spectra as functions of the wall-parallel wavelengths.
\solid, $R'_{12}$; \dashed, optimal $\phi'_{12}$; \chndot, cospectrum of $u'_1$ and $u'_2$. The
three contours contain 10\%, 50\% and 80\% of the spectral mass. The dashed diagonals are
$\lambda_1=\lambda_3$. Flow as in figure \ref{fig:flux3}, $x_2/h=0.15$.
(b) Streamwise premultiplied spectra, as a function of $x_2$. Each horizontal section is a
spectrum normalised to unit energy, and the contours are 50\% and 80\% of the global
maximum.
\solid, $R'_{12}$; \dashed, $\phi'_{12}$. The  dashed diagonal is $\lambda_1=1.2\, x_2$.
}
\label{fig:spectra}
\end{figure} 
   
Reynolds and optimal fluxes are also structurally quite different. This is shown in figures
\ref{fig:flux3}(a,b), where the classical transverse Reynolds stress $R'_{12}\approx u'_1
u'_2$, is compared with the corresponding optimal flux $\phi'_{12}$. Both quantities are
shown centred with respect to their mean and normalised with their standard deviation. This
scaling absorbs the difference in their magnitude, but the geometry of the field remains
different. This is partly because of the stronger intermittency of $R'_{12}$, manifested by
the presence of numerous dark and light spots in figure \ref{fig:flux3}(a), but the
characteristic streamwise organisation of the Reynolds stresses is much less marked in the
optimal flux in figure \ref{fig:flux3}(b).

This is confirmed by the spectra in figure \ref{fig:spectra}, where $\phi'_{12}$ is compared
to $R'_{12}$. Note that these spectra are different from the cospectrum of $u'_1$ and
$u'_2$, which is included in figure \ref{fig:spectra}(a) for comparison. The latter
represents the contribution of the product $\tau_{12}$ to the mean tangential stress, while
the former reflect the geometry of the product \citep{lozano-Q}. Figure \ref{fig:flux3}(a)
shows two-dimensional spectra in the plane $x_2/h=0.15$, and reveals that the cospectrum is
dominated by the elongated streaks of the streamwise velocity. In the case of products of
velocities, the high-order spectra were shown by \cite{VanWyn75} to be dominated by the
sweeping of the small scales by the larger ones, and the spectrum of $R'_{12}$ is also
anisotropic, although less than the cospectrum. On the other hand, the spectrum of
$\phi'_{12}$, which is only indirectly linked to $u_1$ through the right-hand side of
\r{eq:NS2}, is only weakly influenced by the elongated streaks, and is more isotropic
(i.e., closer to $\lambda_1=\lambda_3$).

Figure \ref{fig:spectra}(b) shows one-dimensional pre-multiplied spectra as functions of the
streamwise wavelength and of the distance to the wall. It is known that the wavelength of
the maximum of the cospectrum increases linearly with $x_2$ at high Reynolds numbers
\citep[see figure 1b in][]{jim12_arfm}, but this is still not obvious at the relatively low
Reynolds number of figure \ref{fig:spectra}. The reason is that $R'_{12}$ is dominated by
the effect of the wall-parallel velocity $u'_1$, whose scale is not constrained by the
impermeability condition near the wall \citep{tow:61}. On the other hand, figure
\ref{fig:spectra}(b) shows that the $\phi'_{12}$, which is free from spurious influence of
the inactive wall-parallel motion, grows linearly away from the wall even at this relatively
low Reynolds number. It can be shown that the difference between the spectrum of classical
and optimal fluxes is largest in the buffer layer, and decreases with increasing distance to
the wall. This effect is more marked for quantities involving $u'_1$ or $u'_3$, and almost
nonexistent for $\phi_{22}$.

Note that the gauge freedom in the definition of the momentum fluxes calls into
question the meaning of individual structures of intense Reynolds products, such as those
studied in the classical `quadrant' classification of the $(u'_1,u'_2)$ plane by
\cite{wall:eck:bro:72} and \cite{lu:wil:73}, and in modern three-dimensional extensions of
the same idea by \cite{lozano-Q,lozano-time}. Although a detailed investigation of this
question is beyond the scope of this paper, the present results suggest that these structural
analyses should be repeated using other gauges, such as the present optimal one, to test how
dependent on the gauge are the properties of the resulting structures. This is a case in
which intermittency is beneficial, since the hope is to identify structures strong enough to
stay coherent independently of the rest of the flow, but able to explain some flow
characteristics from a small fraction of the total volume. The weaker intermittency of the
optimal fluxes in figure \ref{fig:interm} suggests that analyses based on intense structures
may be less relevant for them than for the classical Reynolds products. For example, it
follows from the pdfs in figure \ref{fig:interm}(c) that, while the 10\% strongest points of
$-\tau_{12}$ contain around 70\% of the total Reynolds stress, the equivalent strongest 10\%
of the optimal $-\phi'_{12}$ only accounts for 33\%. Correspondingly, the volume fraction of
the `countergradient' momentum flux, defined as $\phi_{12} \p_2U>0$, is approximately 15\%
for the optimal fluxes in the logarithmic layer, and 30\% for $\tau_{12}$.

\section{Discussion and conclusions}\label{sec:conclusions}

We have seen that the fluxes implicit in conservation laws are not uniquely defined, in a
way similar to the gauge ambiguity of classical field theory. As an example, we have
developed a particular definition that minimises the integral of their square. Although this
definition should not be considered in any way unique, it has the intuitive appeal of
generating minimum `sterile circularity' in the transfers of the conserved quantity. We have
presented a way to compute such fluxes from simulations, and applied it to the momentum
transfer in turbulent channels. Of particular interest is that the results differ
substantially from the classical Reynolds stresses, whose main justification appears, in
this light, to be that they have become easier to interpret through familiarity, and that
they are obtained from a particularly convenient algebraic manipulation of the equations of
motion. 
   
The present results raise some interesting questions that go beyond the scope of the present
paper, and which should eventually be considered individually. Not the least of them is
whether, given their arbitrariness, point-wise Reynolds stresses should be considered to be
proper targets for the sub-grid models of large-eddy simulations (LES) or, up to point, of
Reynolds-averaged models. Only their divergence is important, while the stresses themselves
can vary widely without ill effects. This may help explain the apparent contradiction that
`a-priori' testing of many LES models fails grossly while the `a-posteriori' results are
reasonable \citep{bardina}. It is interesting to note in this context that the very
successful dynamic model \citep{germetal01} can be characterised as an algorithm to
determine the magnitude of the eddy viscosity from the difference of the subgrid stresses at
two different scales, and can therefore be seen as an integral implementation of fitting the
divergence of the fluxes in scale space, rather than the fluxes themselves.
 
Also interesting is that \r{eq:divn}--\r{eq:divnR} provides an algorithmic `accounting' definition of
fluxes that can be computed even in cases in which the physical formulation is difficult to
interpret locally. There is no implication that the result can be expressed in terms of a
`formula' of local variables, but this is no worse than for the pressure, which is part of
the classical momentum and energy fluxes, and can only be determined as the solution of a
partial differential equation. Numerically, all variables are equally simple to obtain,
particularly since \r{eq:divnR} ensures that any expression for the fluxes
provides a way to compute all other representations.

For example, the Kolmogorov inertial energy cascade assumes that energy is transferred
across scales from its injection into large structures to its dissipation in small viscous
ones. Defining the energy flux, $\dis$, is straightforward in isotropic flows for which
scale is a one-dimensional parameter, and the conservation equation \r{eq:divn} can be
solved by a simple quadrature. In more general cases the definition is not so clear, and any
attempt to write energy conservation in spectral space leads to a formulation in terms of
wavenumber triads that cannot be interpreted locally. This ambiguity is at the root of many
of the discussions about the instantaneous direction of energy transfer and of the relevance
of backscatter. An energy equation equivalent to \r{eq:divn} in wavenumber space provides a
definition of a vector energy-transfer rate that is local, algorithmically computable and,
inasmuch as energy conservation embodies the behaviour of the energy, as physically
`relevant' as any definition based on algebraic expressions. Note that even the classical
one-dimensional definition of $\dis$ relies on a homogeneous boundary condition such as
\r{eq:varbc}. The energy transfer rate can only be given a definite value by assuming that
it vanishes at very large and very small scales.

An even more interesting application concerns non-homogeneous flows. While the concept of
scale is unambiguous in homogeneous flows, it is harder to define in inhomogeneous ones,
where it is not easily separated from position. One of the central results of harmonic
analysis is that position and scale cannot be exactly defined at the same time. Consider,
for example, a turbulent channel in which energy is being transferred among eddies of
different sizes while they move relative to the wall. The ambiguity is whether their
energy should be considered as being transferred across space or across scale. A recent
analysis of this problem led to an equation for the transfer of the second-order structure
function (the `scale energy') in the form of a double divergence, in space and scale, of a
six-dimensional vector flux \citep{Hill02},
\beq
\p_{x_j} \phi_j+\p_{r_j} \psi_j =S,
\la{eq:div4}
\eeq
where $x_j$ with $j=1\ldots 3$ represent the spatial directions, and $r_j$ are the
respective separations along those directions. The analysis provides explicit expressions
for the fluxes in space, $\phi_j$, and scale, $\psi_j$, which have been computed and
interpreted in turbulent channels by \cite{cimar11} and \cite{cimar12}. They are not optimal
in the sense described above (Cimarelli, private communication). Irrespective of the merits
of the structure function as a measure of energy at a given scale, the previous
considerations show that these fluxes and this analysis are not unique, and suggest that
their conclusions should be revisited in terms of their robustness with regards to the
different definitions.

In general, cascade theories concern themselves with fluxes, which are typically conserved
across some `inertial' range. The quantities being transferred, such as
the energy,  are typically created and dissipated somewhere else in the system. The
results in the present paper suggest that the concept of flux, and therefore of cascades,
should be re-examined with care.
   
\acknowledgments
This work was supported by the European Research Council Coturb grant  ERC-2014.AdG-669505.
I am grateful to the Sidney Sussex College and to the department of Engineering of the U. of Cambridge for their hospitality during the preparation of this manuscript.
%

%

\end{document}